\documentclass{article}
\usepackage{authblk,xspace,hyperref,amssymb,longtable,physics,graphicx,pdflscape}
\usepackage[UKenglish]{babel}
\newcommand{\etal}{\textit{et~al.}\@\xspace}
\newcommand{\SimProp}{\textit{SimProp}\xspace}
\newcommand{\AppendixRef}{Appendix~\ref}
\newcommand{\SectionRef}{Section~\ref}
\newcommand{\TableRef}{Table~\ref}
\newcommand{\mail}[1]{\href{mailto:#1}{\nolinkurl{#1}}}
\newcommand{\eV}{\mathrm{eV}}
\newcommand{\MeV}{\mathrm{MeV}}
\newcommand{\PeV}{\mathrm{PeV}}
\newcommand{\EeV}{\mathrm{EeV}}
\newcommand{\K}{\mathrm{K}}
\newcommand{\km}{\mathrm{km}}
\newcommand{\mb}{\mathrm{mb}}
\newcommand{\Mpc}{\mathrm{Mpc}}
\newcommand{\s}{\mathrm{s}}
\newcommand{\yr}{\mathrm{yr}}
\newcommand{\Einj}{E_\text{inj}}
\newcommand{\zinj}{z_\text{inj}}
\newcommand{\Om}{\Omega_\text{m}}
\newcommand{\OL}{\Omega_\Lambda}
\newcommand{\kB}{k_\text{B}}
\newcommand{\p}{\mathrm{p}}
\newcommand{\n}{\mathrm{n}}
\newcommand{\e}{\mathrm{e}}

\newcommand{\He}{{^4\mathrm{He}}}
\newcommand{\N}{{^{14}\mathrm{N}}}

\title{\SimProp~v2r3: Monte Carlo simulation code of UHECR propagation}
\date{3~February 2016}
\author[a,b]{R.~Aloisio}
\author[c]{D.~Boncioli\footnote{Now at DESY, Zeuthen, Germany.}}
\author[d]{A.~di~Matteo\footnote{Parts of this document are adapted from AdM's PhD thesis at University of L'Aquila}}
\author[c]{A.F.~Grillo}
\author[a,d]{S.~Petrera}
\author[e]{F.~Salamida}
\affil[a]{Gran Sasso Science Institute (INFN), L'Aquila,  Italy}
\affil[b]{INAF/Osservatorio Astrofisico di Arcetri, Firenze, Italy}
\affil[c]{INFN/Laboratori Nazionali Gran Sasso, Assergi, Italy}
\affil[d]{INFN and Department of Physical and Chemical Sciences, University of L'Aquila, L'Aquila, Italy}
\affil[e]{INFN and Department of Physics, University of Milano-Bicocca, Milan, Italy}

\begin{document}
\maketitle
\begin{abstract}
We introduce the new version of \SimProp, a Monte Carlo code for simulating the propagation of ultra-high energy cosmic rays in intergalactic space. This version, \SimProp~v2r3, allows the choice of many more models for the extragalactic background light spectrum and evolution and photodisintegration cross sections and branching ratios than previous versions of \SimProp.
\end{abstract}
\tableofcontents
\section{Motivation and history}

\SimProp is a simple Monte Carlo code for the simulation of the propagation of ultra-high energy cosmic rays in intergalactic space,
originally developed as a refinement of the analytic models by Aloisio, Berezinsky and
Grigorieva~\cite{bib:Aloisio1,bib:Aloisio2} in order to have a publicly available
Monte Carlo code for the community to use~\cite{bib:Denise}, at a time when most UHECR propagation studies
used closed-source simulation codes such as those by Allard \etal~\cite{bib:Allard}.
More sophisticated codes have since become available, such as CRPropa \cite{bib:CRP2,bib:CRP3},
but having a simpler, independently developed code can still be useful for cross-checking the correctness of their results (see e.g.~Ref.~\cite{bib:SALpropa}) in a short computational time.

The processes that \SimProp takes into account are:
   the adiabatic energy loss that all particles undergo due to the expansion of the Universe, \begin{equation}
\qty(-\frac{1}{E}\dv{E}{t})_\text{ad} = H(t) = H_0\sqrt{(1+z)^3\Om+\OL}, \label{eq:adiabatic}
  \end{equation}
  where the values of the cosmological parameters used in \SimProp are $H_0 = 7.11\times 10^{-11}~\yr^{-1}\approx 70~\km/\s/\Mpc$, $\Om = 0.3$, and $\OL = 0.7$;
  the interactions with cosmic microwave background (CMB) and infrared, visible and ultraviolet extragalactic background light (EBL) photons,
  with interaction rate \begin{equation}
\frac{1}{\tau} = \frac{1}{2\Gamma^2}\int_{\epsilon'=0}^{2\Gamma\epsilon}\int_{\epsilon=0}^{+\infty}\frac{n_\gamma(\epsilon)}{\epsilon^2}\dd{\epsilon} \sigma(\epsilon') \epsilon' \dd{\epsilon'} \label{eq:intrate}, 
\end{equation}
  where $\Gamma$~is the Lorentz factor of a particle in the laboratory (lab) frame, $n_\gamma(\epsilon)$~is the number per unit volume per unit energy of background photons with energy~$\epsilon$ in the lab frame, and $\sigma(\epsilon')$~is the cross section for interactions with background photons with energy~$\epsilon'$ in the nucleus rest frame (NRF); 
  and the decays of unstable particles (pions, muons, and beta-decay unstable nuclei), which are treated of instantaneous as decay lengths are generally much shorter than all other relevant length scales.
  
  The photon background used in \SimProp is $n_\gamma = n_\text{CMB} + n_\text{EBL}$, where
  \begin{equation}
n_\text{CMB}(\epsilon,z) = \frac{1}{\pi^2} \frac{\epsilon^2}{\exp(\epsilon/\kB (1+z)T_0)-1}, \label{eq:CMB}
\end{equation}
 where $\kB T_0 = 0.2327~\mathrm{meV}$ ($T_0\approx2.7~\K$), and several models are available for $n_\text{EBL}(\epsilon,z)$.
    The interaction processes implemented in \SimProp are the electron--positron pair photoproduction \begin{equation}
N + \gamma \to N + \e^+ + \e^-
\end{equation}
 approximated as a continuous energy loss with precomputed energy loss rates~\cite{Berezinsky:2002nc}, the photodisintegration of nuclei \begin{align}
{^AZ} + \gamma &\to {^{A-1}Z} + \n, &
{^AZ} + \gamma &\to {^{A-1}(Z-1)} + \p, \\
{^AZ} + \gamma &\to {^{A-4}(Z-2)} + \He, && \text{etc.} \label{eq:alphaejection}
\end{align} for which different models are available,
 and the pion photoproduction \begin{align}
\p + \gamma &\to \p + \pi^0, & \n + \gamma &\to \n + \pi^0, \label{eq:pi0} \\
\p + \gamma &\to \n + \pi^+,\qquad\text{and} & \n + \gamma &\to \p + \pi^-, \label{eq:pi+}
\end{align} with total cross sections computed by SOPHIA~\cite{bib:SOPHIA}, branching ratios computed assuming isospin invariance, distribution of outgoing pion directions taken to be isotropic in the centre-of-mass (CoM) frame, and nuclei treated as collections of free nucleons. Magnetic fields are not taken into account in \SimProp; therefore, the
propagation is taken to be rectilinear, with only one coordinate, the redshift~$z$,
used to keep track of particle positions.

\subsection{Previous versions}
In the first released version, \SimProp~v2r0~\cite{bib:SPv2r0}, the only processes treated stochastically were
the sampling of the source redshift and initial energy of primary particles
and the photodisintegration of nuclei.
All other processes, namely the adiabatic energy loss, pair production, and pion production, were
treated deterministically, as in the analytic models \SimProp was based on.
Pion production was only taken into account for protons interacting with CMB photons and approximated as a continuous energy loss.
Photodisintegration was treated according to the Puget--Stecker--Bredekamp (PSB) model~\cite{bib:PSBsigma}
as refined by Stecker and Salamon~\cite{bib:SSthresholds}, taking into account
the EBL using the Stecker \etal model~\cite{bib:SteckerEBL} or a power-law approximation thereof~\cite{bib:Aloisio1,bib:Aloisio2}
as well as the CMB. 

Starting from the following version, \SimProp~v2r1~\cite{bib:SPv2r1}, the pion production process is also treated stochastically,
in order to compute fluxes of $\EeV$ secondary neutrinos thereby produced and assess the reliability of the continuous energy loss
approximation for that process,  and it affects both protons and other nuclei.

\SimProp~v2r2~\cite{bib:SPv2r2} fixed a few bugs in \SimProp~v2r1, and added an option to take into account pion production on the EBL, in order
to compute secondary neutrino fluxes down to $\PeV$~energies~\cite{Aloisio:2015ega}. It also added the choice of using the Kneiske \etal
model~\cite{bib:KneiskeEBL} for the EBL.

\subsection{Current version}
\SimProp~v2r3 also allows the user to choose the Do\-m\'in\-guez \etal (best fit, lower limit or upper limit)~\cite{bib:DominguezEBL} or the Gilmore \etal~\cite{bib:GilmoreEBL} EBL models, and to use one of four different parametrizations for photodisintegration cross sections with user-defined parameter values. In particular, processes where alpha particles are ejected \eqref{eq:alphaejection} are also implemented, and their cross sections can be scaled by a user-defined factor via a simple command-line option, allowing the user to assess the effects of these poorly known quantities on the results. 

Comparisons between the results from \SimProp~v2r3 and CRPropa~3~\cite{bib:CRP3} with various settings are discussed in Ref.~\cite{bib:SALpropa}, and the effects of their differences on the modelling of UHECR sources are discussed in Ref.~\cite{bib:fitICRC15,bib:CRIS2015}.

\SimProp is available upon request to \mail{SimProp-dev@aquila.infn.it}.

\section{Main features}
A $\SimProp$ run consists of $N$~events. Each event consists of the generation of a primary particle with mass number~$A_\text{inj}$,
 initial energy~$\Einj$ such that $\log_{10}(\Einj/\eV)$~is uniformly distributed from~$l_{\min}$ to~$l_{\max}$, and
source redshift~$\zinj$ uniformly distributed from~$z_{\min}$ to~$z_{\max}$\footnote{This can be converted to different distributions by weighing each event by an appropriate function of $\Einj$ and $\zinj$, for example by $$w(\Einj,\zinj)\propto\frac{\Einj^{1-\gamma}(1+\zinj)^{m-1}}{\sqrt{(1+\zinj)^3\Om+\OL}}$$ for a power-law injection spectrum~$\propto \Einj^{-\gamma}$ and density of sources per unit comoving volume~$\propto (1+\zinj)^m$ (see \AppendixRef{app:distances}).} and its propagation to Earth, along with that of any
secondary particles produced during the propagation; $A_\text{inj}$, $l_{\min}$, $l_{\max}$, $z_{\min}$ and~$z_{\max}$ are input parameters (see below).

The interval between the production of a particle (at injection for primaries, and at an interaction for secondaries) and its decay, stochastic interaction or arrival to Earth is called a branch. During a branch, a particle keeps the same mass number and electric charge, but loses energy through continuous processes, such as the adiabatic energy loss and pair production. Each of the outgoing particles in a stochastic interaction or decay starts a new branch, even if it is of the same type as the parent (e.g.~as in neutral pion production).

\SimProp uses the {\tt TRandom3}~random number generator from the ROOT framework~\cite{bib:ROOT}, which is based on the Mersenne twister~\cite{bib:Mersennetwister}.

\subsection{Input parameters}
\label{sec:input}
\SimProp~v2r3 recognizes the following command-line options:
\begin{longtable}[c]{lp{21em}r}
option & description & default \\
\hline\endhead
{\tt -h} & prints the help and exits & none \\
{\tt -s} & seed of the random number generator & 65539 \\
{\tt -N} & number of events to be generated & 100 \\
{\tt -L} & EBL model: {\tt 0}~none (CMB only); {\tt 1}~Stecker \etal \cite{bib:SteckerEBL}; {\tt 2}~power-law approximation of 1 \cite{bib:Aloisio1,bib:Aloisio2}; {\tt 3}~Kneiske \etal \cite{bib:KneiskeEBL}\footnote{This EBL model is implementing by interpolating the photon density at $z=0$ as a function of $(1+z)\epsilon$ and multiplying it by a scale factor, whereas all other models are interpolated on a 2D grid of $(\epsilon, z)$ values.}; {\tt 4}~Do\-m\'in\-guez \etal \cite{bib:DominguezEBL} best fit; {\tt 5}~Do\-m\'in\-guez \etal lower limit; {\tt 6}~Do\-m\'in\-guez \etal upper limit; {\tt 7}~Gilmore \etal \cite{bib:GilmoreEBL} & 1 \\
{\tt -A} & mass number of primary nuclei, $A_\text{inj}$ (chosen at random for each event with~{\tt -A 0}) & 56 \\
{\tt -S} & treatment of pion production: {\tt -1}~continuous energy loss approximation for protons, neglected for other nuclei (as in \SimProp~v2r0); {\tt 0}~continuous energy loss for both protons and other nuclei; {\tt 1}~stochastic, on the CMB only; {\tt 2}~stochastic, on both the CMB and the EBL & 1 \\
{\tt -D} & beta decay: {\tt 0}~disabled, all nuclei treated as their respective beta-decay stable isobars; {\tt 1}~enabled, treated as instantaneous & 1 \\
{\tt -e} & logarithm of minimum injection energy, $l_{\min}$ & 17 \\
{\tt -E} & logarithm of maximum injection energy, $l_{\max}$ & 21 \\
{\tt -z} & minimum source redshift, $z_{\min}$ & 0 \\
{\tt -Z} & maximum source redshift, $z_{\max}$ & 1 \\
{\tt -r} & distance between sources, $L_\text{s}$,\footnote{If $L_\text{s}$ is nonzero, then whenever an event is generated with $\zinj<0.1$ it is rounded up to the next higher integer multiple of $\Delta z =H_0L_\text{s}=L_\text{s}/(4285~\Mpc)$.} in $\Mpc$ & 0 \\
{\tt -o} & output type (see \SectionRef{sec:output}): {\tt 0}~old ({\tt nuc} and {\tt ev} trees); {\tt 1}~new ({\tt summary} tree); {\tt 2}~both & 0 \\
{\tt -M} & photodisintegration model: {\tt 0}~PSB~\cite{bib:PSBsigma} with Stecker--Salamon thresholds~\cite{bib:SSthresholds}; {\tt 1}~arbitrary Gaussians (see below); {\tt 2}~arbitrary Breit--Wigner functions; {\tt 3}~arbitrary Breit--Wigner functions with alpha particle ejection; {\tt 4}~arbitrary Gaussians with alpha particle ejection & 0 \\
{\tt -n} & nucleon ejection scaling factor (only with {\tt -M~3} and {\tt -M~4}) & 1 \\
{\tt -a} & alpha-particle ejection scaling factor (only with {\tt -M~3} and {\tt -M~4}) & 1 \\
\end{longtable}

If {\tt -M}~is used with a nonzero value, the parameters for the photodisintegration model are read from the standard input as follows: 
the first line must contain~$n$, $\epsilon_1/\MeV$, and $\epsilon_{\max}/\MeV$, where $n$~is the number of different nuclides to be treated (excluding free nucleons; 51 in the PSB model), and the meanings of~$\epsilon_1$ and~$\epsilon_{\max}$ depend on the model. Afterwards, there must be one line for each of the nuclides to be treated, in decreasing order of $A$, giving a series of parameters, as follows:

\paragraph{Photodisintegration model 1: arbitrary Gaussians}
This model has the same structure as the PSB model, described in \AppendixRef{sec:disimodels}, but the values of the parameters are read from the input; the meanings of the parameters are the same as described in \AppendixRef{sec:disimodels}. The entry for each nuclide must give $Z$, $A$, $\epsilon_{\min,1}$, $\epsilon_{\min,2}$, $\epsilon_{0,1}$, $\xi_1$, $\Delta_1$, $\epsilon_{0,2}$, $\xi_2$, $\Delta_2$, and $\zeta$, in this order. All energies must be given in~$\MeV$.

\paragraph{Photodisintegration model 2: arbitrary Breit--Wigner functions}
This model includes the same processes as the previous one, but for $A>4$ the shape of the cross sections for the one- and two-nucleon ejection processes are replaced by
\begin{align}
\sigma_i(\epsilon') &= \frac{\xi_i}{1+{(\epsilon'-\epsilon_{0i})^2}/{\Delta_i^2}}, \quad i=1,2.
\end{align}
For $A\le4$, PSB cross sections are used.

\paragraph{Photodisintegration model 3: arbitrary Breit--Wigner functions with alpha-particle ejection}
In this model, two processes are treated: single-nucleon ejection and alpha-particle ejection.
The cross sections used are of the form:
\begin{align}
  \sigma_{N}(\epsilon') &=
  \begin{cases}
    \frac{h_{N1}}{1+\qty(\frac{\epsilon'-x_{N1}}{w_{N1}})^2} + \frac{h_{N2}}{1+\qty(\frac{\epsilon'-x_{N2}}{w_{N2}})^2}, & t_N < \epsilon' \le \epsilon_1;\\
    c_N, & \epsilon_1 < \epsilon' \le \epsilon_{\max};\\
    0, & \text{otherwise};
  \end{cases} \\
  \sigma_\alpha(\epsilon') &=
  \begin{cases}
    \frac{h_{\alpha1}}{1+\qty(\frac{\epsilon'-x_{\alpha1}}{w_{\alpha1}})^2} + \frac{h_{\alpha2}}{1+\qty(\frac{\epsilon'-x_{\alpha2}}{w_{\alpha2}})^2}, & t_\alpha < \epsilon' \le \epsilon_1;\\
    c_\alpha, & \epsilon_1 < \epsilon' \le \epsilon_{\max};\\
    0, & \text{otherwise}.
  \end{cases} 
\end{align}
For nuclei with $A \le 4$, PSB cross sections~$\sigma_1, \sigma_2, \sigma_3$ are computed, and then $\sigma_{N}=\sigma_1 + 2 \sigma_2 + 1.2 \sigma_3, \sigma_\alpha=0$~are used. If the command-line options {\tt -n} and/or {\tt -a} are used, all values of~$\sigma_N$ and~$\sigma_\alpha$ used are scaled by the corresponding parameters.

The $\SimProp$~v2r3 package contains two files intended for use with this model, \nolinkurl{talys10sigma} and \nolinkurl{talys16sigma}, where the values of the parameters were fitted via eqs.~(\ref{eq:nucl}, \ref{eq:alph}) to cross sections computed by TALYS-1.0~\cite{bib:TALYS} and TALYS-1.6 respectively with their default settings.

\paragraph{Photodisintegration model 4: arbitrary Gaussians with alpha-par\-ti\-cle ejection}
This model treats the same processes as the previous one, but the cross sections used are:
\begin{align}
  \sigma_{N}(\epsilon') &=
  \begin{cases}
    h_{N1}\exp(-\frac{(\epsilon' - x_{N1})^2}{w_{N1}}), & t_N < \epsilon' \le \epsilon_1;\\
    c_N, & \epsilon_1 < \epsilon' \le \epsilon_{\max}; \\
    0, & \text{otherwise};
  \end{cases} \\
  \sigma_\alpha(\epsilon') &=
  \begin{cases}
    h_{\alpha1}\exp(-\frac{(\epsilon' - x_{\alpha1})^2}{w_{\alpha1}}), & t_\alpha < \epsilon' \le \epsilon_1;\\
    c_\alpha, & \epsilon_1 < \epsilon' \le \epsilon_{\max};\\
    0, & \text{otherwise}.
  \end{cases} 
\end{align}
For nuclei with $A \le 4$, PSB cross sections~$\sigma_1, \sigma_2, \sigma_3$ are computed, and then $\sigma_N=\sigma_1 + 2 \sigma_2 + 1.2 \sigma_3, \sigma_\alpha=0$~are used. If the command-line options {\tt -n} and/or {\tt -a} are used, all values of~$\sigma_N$ and~$\sigma_\alpha$ used are scaled by the corresponding parameters.

The $\SimProp$~v2r3 package contains a file intended for use with this model, \nolinkurl{pars_talysfixed.txt}, where the values of the parameters were fitted via eqs.~(\ref{eq:nucl}, \ref{eq:alph}) to cross sections computed by TALYS-1.6 with settings restored to those used in Ref.~\cite{bib:Khan} (see \AppendixRef{sec:disimodels}).

\subsection{Output files}\label{sec:output}
\SimProp~v2r3 writes its output in a ROOT~\cite{bib:ROOT} file whose name encodes the command-line options used (e.g.~\nolinkurl{SimProp-v2r3_N100_A56_L1_S1_D1_z0.00_Z1.00_e17.0_E21.0_Ls0.00_M0_n1.00_a1.00_o0_s65539.root} when the default parameter values are used).

If the parameter~{\tt -o} is set to 0, the file contains the following trees:
\begin{description}
\item[Tree {\tt nuc}] This tree has an entry for each branch. The name of the tree is due to historical reasons; neutrinos and photons are also included now.
\begin{longtable}[c]{lp{21em}}
branch & description \\
\hline\endhead
{\tt evt~~~~~~~~~~~~} & event number (starting from 0) \\
{\tt branch} & branch generation number: $0$~for the primary, incremented by~$1$ from the parent branch in stochastic interactions and decays\\
{\tt intmult} & $0$~if the particle reaches Earth; $n$~if the particle stochastically interacts producing $n$~secondaries; $1000$ for photons, whose propagation is not yet implemented; $1000+n$~if the particle decays into $n$~particles \\
{\tt Acurr} & mass number during the branch (0 for neutrinos and photons) \\
{\tt Zecurr} & electric charge during the branch \\
{\tt Flav} & flavours of neutrinos ($+1$~for~$\nu_\e$, $-1$~for~$\bar\nu_\e$, $+2$~for~$\nu_\mu$, $-2$~for~$\bar\nu_\mu$), $0$~for all other particles\\
{\tt zOri} & redshift at the beginning of the branch \\
{\tt zEnd} & redshift at the end of the branch \\
{\tt EOri} & energy at the beginning of the branch, in~$\eV$ \\
{\tt EEnd} & energy at the end of the branch, in~$\eV$ \\
{\tt Dist} & comoving distance travelled (see \AppendixRef{app:distances}), in~$\Mpc$ \\
\end{longtable}
\item[Tree {\tt ev}] This tree has an entry for each event.
\begin{longtable}[c]{lp{21em}}
branch & description \\
\hline\endhead
{\tt timexev~~~~~~~~} & CPU time used during the event, in seconds \\
{\tt branxev} & total number of branches in the event \\
{\tt seed} & seed of the random number generator at the end of the event \\
\end{longtable}
\end{description}

If the parameter~{\tt -o} is set to 1, the file contains the following tree:
\begin{description}
\item[Tree {\tt summary}] This tree has an entry for each event.
\begin{longtable}[c]{lp{21em}}
branch & description \\
\hline\endhead
{\tt event} & event number (starting from 0) \\
{\tt injEnergy} & injection energy, in $\eV$ \\
{\tt injRedshift} & source redshift \\
{\tt injDist} & source comoving distance (see \AppendixRef{app:distances}), in~$\Mpc$ \\
{\tt injA} & injection mass number \\
{\tt injZ} & injection atomic number \\
{\tt nNuc} & number of nuclei reaching Earth \\
{\tt nucEnergy[nNuc]} & energies of nuclei reaching Earth, in $\eV$ \\
{\tt nucA[nNuc]} & mass numbers of nuclei reaching Earth \\
{\tt nucZ[nNuc]} & atomic numbers of nuclei reaching Earth \\
{\tt nPho} & number of photons produced from $\pi^0$ decay \\
{\tt phoEProd[nPho]} & energies of photons at production, in $\eV$ \\
{\tt phozProd[nPho]} & redshifts of production points of photons \\
{\tt nNeu} & number of neutrinos reaching Earth \\
{\tt neuEnergy[nNeu]} & energies of neutrinos reaching Earth, in $\eV$ \\
{\tt neuFlav[nNeu]} & flavours of neutrinos reaching Earth ($+1$~for~$\nu_\e$, $-1$~for~$\bar\nu_\e$,
$+2$~for~$\nu_\mu$, $-2$~for~$\bar\nu_\mu$) \\
\end{longtable}
\end{description}

If the parameter~{\tt -o} is set to 2, the file contains all the trees described above.

\section{Structure of the code}
\SimProp is written in C++, and makes use of a few features from the ROOT~\cite{bib:ROOT} framework.

During each event, a stack contains all the particles to be propagated. At the beginning of the event, the stack only contains the primary particle, with user-specified mass number and with initial energy and redshift sampled from the user-specified ranges. New particles are added to the stack when produced during the propagation, and particles that decay, stochastically interact or reach Earth are removed from it. The event is over when the stack becomes empty.

\subsection{Propagation of protons and stable nuclei}
When propagating a proton or a stable nucleus, the redshift interval between its production point $z_\text{prod}$ and $0$ (Earth) is divided into steps $z_0 = z_\text{prod}$, $z_1$, \ldots, $z_n = 0$, shorter near the production point than near Earth. During each step, there are two types of processes the particle can undergo:
\begin{itemize}
\item those which are treated as continuous (deterministic) energy losses and do not involve the production of any new particles to be tracked, namely the adiabatic energy loss, pair production on CMB photons, and (if the option {\tt -S~-1} or {\tt -S~0} is used) pion production on CMB photons;
\item those which are treated as discrete interactions, with the interaction point and the energies, types and/or number of outgoing particles to be sampled stochastically, namely the photodisintegration of nuclei and pion production (on CMB photons by default or if the option {\tt -S~1} is used, and also on EBL photons if {\tt -S~2} is used).
\end{itemize}

\subsubsection{Continuous energy losses}
At each redshift step $(z_{i-1},z_i]$,  the continuous energy losses are simply treated by numerically integrating from $z_{i-1}$ to $z_{i}$ the differential equation for $\ln \Gamma$
\begin{equation} 
\dv{\ln \Gamma}{z} = -\beta(Z,A,\Gamma,z) \dv{t}{z}
\label{eq:cont} 
\end{equation}
where $\Gamma$ is the Lorentz factor of the particle, $\beta$ is the fractional energy loss per unit time, and $-\dd{t}/\dd{z}$ is given by \begin{equation}
\dv{t}{z} = -\frac{1}{(1+z)H(t)} = -\frac{1}{H_0(1+z)\sqrt{(1+z)^3\Om+\OL}}.
\end{equation}
\SimProp uses $\Om = 0.3$, $\OL = 0.7$, and $H_0 = 7.11\times 10^{-11}~\yr^{-1}\approx 70~\km/\s/\Mpc$. The function $\beta$ in eq.~\eqref{eq:cont} is the sum of two terms, one for the redshift loss and one for electron-positron pair photoproduction. The former is computed via eq.~\eqref{eq:adiabatic}
 and the latter as
\begin{equation}
\beta_\text{pair}(Z,A,\Gamma,z) = \frac{Z^2}{A}(1+z)^3\beta_\text{pair}(\text{proton},(1+z)\Gamma,z=0),
\end{equation}
where $\beta_\text{pair}$~for protons at~$z=0$ is interpolated from a list of tabulated values, computed as described in Ref.~\cite{Berezinsky:2002nc}. For non-proton nuclei, $Z^2/A$~is approximated as~$A/4$ (exact when~$A = 2Z$).

\subsubsection{Discrete interactions}
The following scheme is used to decide whether and when the particle undergoes a discrete interaction and, if it does, the type and the products of the interaction.

\paragraph{Sampling of the interaction point}
At the beginning of the propagation, a random number $u$ is sampled from the uniform distribution between $0$ and $1$. The probability that the particle survives to redshift $z$ without interactions is given by 
\begin{equation}
-\ln p = \int_{z_\text{prod}}^{z} \frac{1}{\tau} \dv{t}{z} \dd{z},
\end{equation} 
where the total interaction rate (probability per unit time) $1/\tau$ is computed at each step $z_i$ as described below, and the integral is approximated via the trapezoidal rule, i.e.,
\begin{equation}
-\ln p_i = -\ln p_{i-1} + \frac{1}{2}\left( \left. \frac{1}{\tau} \dv{t}{z} \right|_{z_{i-1}} +  \left. \frac{1}{\tau} \dv{t}{z} \right|_{z_{i}}  \right)(z_{i}-z_{i-1}).
\end{equation}

If at the end of a step $p_i < u$, the particle is considered to have interacted during that step; the interaction point~$z_\text{int}$ is found by linearly interpolating $p$ between~$z_{i-1}$ and~$z_i$ and solving for~$p(z_\text{int}) = u$, and the interaction energy~$E_\text{int}$ is found by integrating eq.~\eqref{eq:cont} from~$z_{i-1}$ to~$z_\text{int}$. These are used to sample the number, type and energy of the outgoing particles as described below, adding these particles to the stack.

If at the end of the last step $p_n > u$, the particle is considered to have reached Earth; its mass number, atomic number, and final energy are recorded in the output file.

\paragraph{Interaction rate}
The total interaction rate~$\tau^{-1}$ is given by eq.~\eqref{eq:intrate}. We compute it as the sum of a term for pion production on the CMB $\tau_\text{pion,CMB}^{-1}$, one for pion production on the EBL $\tau_\text{pion,EBL}^{-1}$, and one for photodisintegration $\tau_\text{disi}^{-1}$. We assume that a nucleus behaves as $A$ independent nucleons in pion production, i.e., $\tau_\text{pion}^{-1}(A,\Gamma,z) = A\tau_\text{pion}^{-1}(\text{proton},\Gamma,z)$, because the energies involved are much larger than the binding energy per nucleon.

We introduce the quantities
\begin{align} 
  I(\epsilon) &= \int_\epsilon^{+\infty}  \frac{n_\gamma(\varepsilon)}{2\varepsilon^2} d\varepsilon; \label{eq:I}\\ 
  \Phi(s) &= \int_{s_{\min}}^{s} (s'-m^2)\sigma(s')\,ds'
          = 4m^2\int_{\epsilon'_{\min}}^{\epsilon'} \varepsilon'\sigma(\varepsilon')\,d\varepsilon' \label{eq:Phi},
\end{align} where $m$ is the mass of the particle (a nucleus in the case of disintegration and a nucleon in the case of pion production) and $s$ is the CoM energy squared $s=m^2+2m\epsilon'$; eq.~\eqref{eq:intrate} can be also written as 
\begin{align}
\frac{1}{\tau} &= \frac{1}{4m^2\Gamma^2} \int_{{\epsilon'_{\min}}/{2\Gamma}}^{+\infty} \Phi(m^2+4m\Gamma\epsilon) \frac{n_\gamma(\epsilon)}{2\epsilon^2} \dd{\epsilon} \label{tau1} \\
~ &= \frac{1}{\Gamma^2} \int_{\epsilon'_{\min}}^{+\infty} I \left(\frac{\epsilon'}{2\Gamma}\right) \epsilon'\sigma(\epsilon') \dd{\epsilon'}. \label{tau2}
\end{align}
The photon background $n_{\gamma}$ is the sum of two terms, one for the CMB and one for the EBL. The CMB spectrum is precisely known at all redshifts~\eqref{eq:CMB}.
 We have
\begin{equation}
I_\text{CMB}(\epsilon) = -\frac{\kB T}{2\pi^2}\ln(1-\exp(-\frac{\epsilon}{\kB T})).
\end{equation}
this implies that $\tau_\text{pion,CMB}^{-1}(\Gamma,z) = (1+z)^3 \tau^{-1}_\text{pion,CMB}((1+z)\Gamma,z=0)$. On the other hand, EBL is not precisely known, and needs to be approximated using phenomenological models. The EBL models available in \SimProp are listed in \SectionRef{sec:input}.

As for the cross sections, we used SOPHIA~\cite{bib:SOPHIA} to compute the pion photoproduction cross section~$\sigma_\text{pion}(\epsilon')$ for protons, and numerically integrated eq.~\eqref{eq:Phi} to obtain a table of values from which we interpolate $\Phi_\text{pion}(s)$. As for photodisintegration, various models are available, listed in \SectionRef{sec:input}.

Finally, $\tau_\text{pion,CMB}^{-1}(A,\Gamma,z)$ is computed by
\begin{equation}
\tau_\text{pion,CMB}^{-1}(A,\Gamma,z)=(1+z)^3A \tau^{-1}_\text{pion,CMB}(\text{proton},(1+z)\Gamma,z=0),
\end{equation} where $\tau^{-1}_\text{pion,CMB}$~for protons at~$z=0$ is interpolated from a table whose values were obtained by numerically integrating eq.~\eqref{tau1}; $\tau^{-1}_\text{pion,EBL}$ is computed as a function of $\Gamma$ and $z$ via 2D interpolation from a table obtained by numerically integrating eq.~\eqref{tau2}; and $\tau^{-1}_\text{disi}$ is computed by numerically integrating eq.~\eqref{tau2} when needed.

When a particle interacts, we sample the type of interaction, the probability of each type being $p_j = \tau^{-1}_{j} / \tau^{-1}_\text{tot}$. If the interaction is photodisintegration, one of the channels ($\sigma_1$, $\sigma_2$ and $\sigma_3$ with options {\tt -M~0}~to {\tt -M~2}, and $\sigma_N$ and $\sigma_\alpha$ with {\tt -M~3}~and {\tt -M~4}) is similarly selected and, if $\sigma_3$~is selected, the number of nucleons ejected is sampled from \TableRef{tab:PSBbranch}.

\paragraph{Sampling the number, type and energy of secondary particles}
\subparagraph{Photodisintegration.}
When a nucleus is photodisintegrated, its energy is assumed to be split among the residual nucleus and the ejected fragments in proportion to their mass, i.e.~all the fragments inherit the Lorentz factor of the original nucleus.  It is assumed that each nucleon has the same probability of being ejected, regardless of its type (i.e., if a nucleus with 26 protons and 30 neutrons loses a nucleon, it is assumed to be a proton with probability~$26/56$ and a neutron with probability~$30/56$); this simplifying assumption is only approximately realistic (as in reality interaction channels yielding stable nuclei are more likely) and may result in a slight overestimate of the number of beta decays (and resulting neutrinos).

In the photodisintegration models {\tt -M~0} to {\tt -M~2}, the processes considered are single nucleon ejection, double nucleon ejection, and multiple nucleon ejection; the type of each nucleon ejected is sampled at random.

With photodisintegration models {\tt -M~3} and {\tt -M~4}, the processes implemented are single nucleon ejection and alpha-particle ejection. The rationale for not considering multiple nucleon ejection in these models is that, while the statistical uncertainties associated with the available $\ln A$ measurements
    are in principle small enough to distinguish protons from helium-4, they
    are too large to distinguish consecutive intermediate nuclei, e.g. carbon-12
    from carbon-13. Therefore it is important that UHECR propagation simulations
    accurately predict the number of protons and alpha particles reaching Earth,
    but it is unnecessary to have the correct distribution of individual intermediate masses.
    The interaction rates (and hence the cross sections) for these processes can be taken to be the sum of those for all actual processes weighted
    by the number of nucleons and alpha particles ejected, respectively.
    This ensures that the numbers of free nucleons and of alpha particles at Earth, assuming
    that the interaction rates do not change too rapidly with $z$ or $A$, are reproduced in good approximation.\footnote{
        For example, assume we have $\N$~nuclei originating 70~Mpc away, and the
        only relevant process is~$\N + \gamma \to ^{12}\mathrm{C} + \p + \n$,
        with interaction length 100~Mpc. A fraction~$\exp(-0.7) \approx 50\%$ of the nuclei will survive,
        and at Earth, for each 100 $\N$~nuclei injected, we will have
        in average 50 $\N$~nuclei, 50 $^{12}\mathrm{C}$~nuclei, and
        100 free nucleons. If we chose to approximate this process as~$\N + \gamma \to ^{13}\mathrm{C} + \mathrm{p}$
        and~$^{13}\mathrm{C} + \gamma \to ^{12}\mathrm{C} + \n$ with interaction
        length 50~Mpc each, a fraction~$\exp(-0.7)^2 \approx 25\%$ of the nuclei will survive,
        $2\exp(-0.7)(1-\exp(-0.7)) \approx 50\%$ will interact once, and $(1-\exp(-0.7))^2 \approx 25\%$ will interact twice,
        and at Earth, for each 100 $\N$~nuclei injected, we will have
        in average 25 $\N$~nuclei, 50 $^{13}\mathrm{C}$ nuclei, 25 $^{12}\mathrm{C}$ nuclei, and
        100 free nucleons. Both the number of free nucleons and the average mass of the intermediate nuclei will then be
        well approximated, though the numbers of individual intermediate nuclides will be different.
    }
    
    Since deuterium, tritium and helium-3 have very short disintegration lengths,
    such ejectiles can be treated as collections of free nucleons; also, since neutrons
    have a short decay length except at extremely high energy and even then the
    air showers they produce are indistinguishable from those of protons,
    all nucleons can be treated as the same, and the only inaccuracies that this approximation can introduce
    are in the fluxes of neutrinos from beta decay, which in any event are strongly subdominant with respect
    to those from pion decay except at the lowest  energies.
    
    Therefore, the cross sections~$\sigma_N$ and~$\sigma_\alpha$ for these two processes can be taken as
    \begin{align}
        \sigma_N &= \sum_\text{channels} n_N \sigma_{n_\n n_\p n_\mathrm{d} n_\mathrm{t} n_\mathrm{h} n_\alpha} =
        \expval{n_N} \sigma_\text{tot}; \label{eq:nucl} \\
        \sigma_\alpha &= \sum_\text{channels} n_\alpha \sigma_{n_\n n_\p n_\mathrm{d} n_\mathrm{t} n_\mathrm{h} n_\alpha}
        = \expval{n_\alpha}\sigma_\text{tot}, \label{eq:alph}
    \end{align}
    where $\sigma_{n_\n n_\p n_\mathrm{d}n_\mathrm{t}n_\mathrm{h}n_\alpha}$ is the
    exclusive cross section for the photodisintegration channel ejecting $n_\n$ neutrons,
    $n_\p$ protons, \ldots, and $n_\alpha$ alpha particles, and  $n_N = n_\n + n_\p + 2n_\mathrm{d} + 3n_\mathrm{t} + 3n_\mathrm{h}$. 
    The files  \nolinkurl{talys10sigma}, \nolinkurl{talys16sigma} and \nolinkurl{pars_talysfixed.txt} provided in the \SimProp~v2r3 package
    contain parameters of fits to~$\sigma_N$ and~$\sigma_\alpha$ defined in this way from the $\sigma_{n_\n n_\p n_\mathrm{d}n_\mathrm{t}n_\mathrm{h}n_\alpha}$ computed by various versions of TALYS (see \SectionRef{sec:input} for more information).

\subparagraph{Pion photoproduction.}
When a pion is photoproduced, if the incoming particle is a nucleus, the nucleon that undergoes the interaction is chosen at random. We approximate all photo-hadronic processes as single-pion production; assuming isospin invariance, a neutral pion is produced~\eqref{eq:pi0} with probability~$1/3$ and a charged pion is produced~\eqref{eq:pi+} with probability~$2/3$.

In order to sample the pion energy, first the photon energy $\epsilon$ in the lab frame is sampled from its marginal distribution\footnote{In practice, we use the fact that the marginal distribution of $\epsilon$ corresponds to a distribution of $I(\epsilon)$ proportional to $\Phi(m^2+4m\Gamma\epsilon)$, and the conditional distribution of $s$ given $\epsilon$ corresponds to a uniform distribution of $\Phi(s)$, so we actually sample $I$ and $\Phi$ and invert the functions to find the corresponding $\epsilon$ and $s$.}  
\begin{align}
p(\epsilon)\dd{\epsilon} &= \frac{\tau}{4m^2\Gamma^2}\Phi(m^2+4m\Gamma\epsilon)\frac{n_{\gamma}(\epsilon)}{2\epsilon^2}\dd{\epsilon}, \quad \frac{\epsilon'_{\min}}{2\Gamma} <\epsilon < +\infty,
\end{align} 
where $m$~is the nucleon mass and~$\epsilon'_{\min} = m_{\pi} + m_{\pi}^2/2m$; then  the squared CoM energy~$s$ is sampled from its conditional distribution given~$\epsilon$
\begin{align}
p(s|\epsilon)\dd{s} &= \frac{(s-m^2)\sigma(s)\dd{s}}{\Phi(m^2+4m\Gamma\epsilon)}, \quad (m+m_{\pi})^2 < s < m^2+4m\Gamma\epsilon~,
\end{align} from~$s$  the pion energy and momentum in the CoM frame are calculated as  
\begin{align}
E^*_{\pi} &= \frac{s - m^2 + m_{\pi}^2}{2\sqrt{s}}; & p^*_{\pi} &= \frac{\sqrt{\left(s-(m+m_{\pi})^2\right)\left(s-(m-m_{\pi})^2\right)}}{2\sqrt{s}}
\end{align}
and the Lorentz factor of the transformation from the CoM frame to the lab frame as $\gamma = m\Gamma/\sqrt{s}$; then the pion energy is converted to the lab frame as~$E_{\pi} = \gamma(E^*_{\pi}+p^*_{\pi}\cos\theta_{\pi})$, where the distribution of~$\theta_{\pi}$ is approximated as isotropic ($\cos\theta_{\pi}$ uniformly distributed between $-1$ and $1$). In the lab frame, the momentum component orthogonal to the original travel direction is much smaller than that parallel to it, by a factor of order $\epsilon/E \sim 10^{-20}$, so we neglect transverse components continuing to assume one dimensional propagation.

The pion with energy $E_\pi$, the nucleon with energy $m\Gamma - E_\pi$, and (in the case of nuclei) a nucleus with mass number $A-1$ and energy $(A-1)m\Gamma$ are then added to the stack.

\subsection{Decay of unstable particles}\label{sec:dec}
When an unstable particle is produced, it is assumed to decay instantaneously, as decay lengths are generally much shorter than all other relevant length scales. The energies of the decay products are sampled as described below and the decay products are added to the stack.

\paragraph{Beta decay of neutrons and unstable nuclei}
Neutrons and nuclei not in the list of beta-decay stable isobars are assumed to immediately undergo beta decay. The $Q$-value of the reaction is read from a table taken from Ref.~\cite{bib:masses} or, for nuclei not on that table, estimated via the  semi-empirical mass formula.

Then, the electron energy in the NRF $E^*_\e$ is sampled from a distribution $\propto (E^{*2}_\e-m_\e^2)^{1/2} E^*_\e (Q - (E^*_\e - m_\e))^2$ (i.e., neglecting electromagnetic effects) and the neutrino energy is calculated as $E^*_{\nu} = Q - (E^*_\e - m_\e)$; the recoil of the nucleus is neglected. The neutrino energy is converted to the lab frame by $E_{\nu} = \Gamma E^*_{\nu} (1 - \cos\theta)$, where $\Gamma$ is the Lorentz factor of the nucleus and $1-\cos\theta$ is sampled from the uniform distribution between $0$ and~$2$. 

The daughter nucleus (with the same energy and mass number $A$ as the parent, with electric charge $Z$ incremented in $\beta^-$ decay and decremented in $\beta^+$~decay) and the neutrino ($\bar{\nu}_e$ in~$\beta^-$ decay, $\nu_e$ in~$\beta^+$ decay) are then added to the stack.

\paragraph{Neutral pion decay}
A $\pi^0$ with energy~$E_\pi$ decays into two photons with energy $E_{\gamma_1}$ distributed uniformly from~0 to~$E_\pi$ and $E_{\gamma_2} = E_\pi - E_{\gamma_1}$.

\paragraph{Charged pion decay}
A $\pi^\pm$ with energy~$E_\pi$ decays into a muon with energy~$E_\mu$ distributed uniformly from~$0$ to~$ (1-m_\mu^2/m_\pi^2)E_\pi$ and a neutrino with energy $E_\nu = E_\pi - E_\mu$.

\paragraph{Muon decay}
A muon with energy $E_\mu$ decays into two neutrinos and an electron (ignored in {\it SimProp}); the energies~$E_{\nu_1}, E_{\nu_2}$ of the neutrinos are sampled as follows: \begin{itemize}
\newcommand{\pone}{p^*_{\nu_1}}
\newcommand{\pones}{p^{*2}_{\nu_1}}
\newcommand{\ptwo}{p^*_{\nu_2}}
\newcommand{\ptwos}{p^{*2}_{\nu_2}}
\newcommand{\pthr}{p^*_\mathrm{e}}
\newcommand{\pthrs}{p^{*2}_\mathrm{e}}
\item the energies of the neutrinos in the muon rest frame $E^*_{\nu_1}$ and $E^*_{\nu_2}$   are sampled independently uniformly from 0 to $m_\mu/2 - m_\e^2/2m_\mu$, and that of the electron is $E^*_\e = m_\mu - E^*_{\nu_1} - E^*_{\nu_2}$;
\item the corresponding momenta are computed as $p^*_{\nu_1} = E^*_{\nu_1}$, $p^*_{\nu_2} = E^*_{\nu_2}$, and $p^*_\mathrm{e} = \sqrt{E^{*2}_\mathrm{e}-m_\mathrm{e}^2}$;
\item if these values violate any of the constraints $E^*_\mathrm{e} \ge m_\mathrm{e}$, $\pone \le \ptwo + \pthr$, $\ptwo \le \pthr + \pone$, or $\pthr \le \pone + \ptwo$, they are discarded and a new $E^*_{\nu_1}, E^*_{\nu_2}$ pair is sampled;
\item the angle $\theta_{12}$ between the two neutrinos is given by \begin{equation}\cos\theta_{12} = \frac{\pthrs-\pones-\ptwos}{2\pone\ptwo};\end{equation}
\item the angle $\theta_1$ between the first neutrino and the line of sight is isotropic, i.e.~$\cos\theta_1$ uniform from $-1$ to $1$;
\item the angle $\phi$ between the second neutrino and the plane containing the line of sight and the first neutrino is uniform from 0 to $2\pi$;
\item the angle $\theta_2$ between the second neutrino and the line of sight is given by $\cos\theta_2 = \cos\theta_{12}\cos\theta_1 - \sin\theta_{12}\sin\theta_1\cos\phi$;
\item finally, the neutrino energies are transformed to the lab frame via $E_{\nu_1} = \gamma(E^*_{\nu_1}+p^*_{\nu_1}\cos\theta_1)$ and $E_{\nu_2} = \gamma(E^*_{\nu_2}+p^*_{\nu_2}\cos\theta_2)$.
\end{itemize}

\subsection{Other particles: photons, electrons and neutrinos}
The propagation of photons and electrons produced is not yet implemented in {\it SimProp}: photons have their production energy and redshift recorded in the output file and electrons are disregarded altogether. The propagation of neutrinos is trivial: no interaction is possible (flavour oscillations are not implemented) and the only energy loss is the redshift loss, so a neutrino produced with energy~$E_\text{prod}$ at redshift~$z_\text{prod}$ will reach Earth with energy~$E_\text{Earth}=E_\text{prod}/(1+z_\text{prod})$.

\section{Future directions}
The next feature we are planning to implement in \SimProp is the computation of fluxes of secondary gamma rays produced in electromagnetic cascades initiated by electrons produced by pair production, muon decay and beta decay and photons produced by neutral pion decay. As discussed in Ref.~\cite{bib:secondarygamma}, the eventual shape of the spectrum of secondary gamma rays is independent on the energy of the primary electrons or gamma rays, so only one quantity, the total energy in these cascades $\Omega_\text{cas}$, will need to be computed.

Another study we are interested in is that of magnetic deflections and the anisotropy in UHECR arrival directions. We expect that under certain reasonable approximations, this will not require any changes to the \SimProp code, but only the assignment of a source position $(z_\text{inj}, \alpha_\text{inj}, \delta_\text{inj})$ to each event during the data analysis, and the simulation of a magnetic deflection from $(\alpha_\text{inj}, \delta_\text{inj})$ to the observed arrival direction $(\alpha_\text{obs}, \delta_\text{obs})$ depending on the source redshift~$z_\text{inj}$ and the initial and final magnetic rigidities.

\appendix
\section{Distance measures}\label{app:distances}
In an expanding universe, there are several possible definitions of distance which for sizeable~$z$ are not equivalent. The following definitions are valid for flat space ($\Omega = \Om+\OL = 1$).

The comoving distance is the proper distance between the positions of two objects measured at a fixed time, divided by the scale factor~$a(t)=R(t)/R_0=(1+z)^{-1}$ at that time. The comoving distance of two objects moving with the Hubble flow does not vary with time. The comoving distance of an object whose light reaches us today after leaving the object at redshift~$z$ is given by \begin{equation}d_\text{C}(z)= \int_0^z \frac{\dd{z}}{H_0\sqrt{(1+z)^3\Om+\OL}}.\end{equation} Likewise, the comoving volume is given by the proper volume times~$(1+z)^3$; the number density per unit comoving volume of a fixed number of objects moving with the Hubble flow does not vary with time.

The light travel distance is the cosmological time elapsed since the light leaves an object until it reaches us. It is given by \begin{equation}d_\text{T}(z)=\int_0^z \abs{\dv{t}{z}} \dd{z} = \int_0^z \frac{\dd{z}}{H_0(1+z)\sqrt{(1+z)^3\Om+\OL}}.\end{equation}

In the case of a distribution of closely spaced identical sources, we can define the source emissivity~$\mathcal{L}$ as the total energy injected per unit comoving volume per unit time, i.e.~$\mathcal{L} = n_\text{s} L$, where $n_\text{s}$~is the number density of sources per unit comoving volume, and the luminosity~$L$ of each source is the total energy emitted by the source per unit time,
\begin{equation}
  L = \int_{0}^{+\infty} E Q(E_\text{inj}) \dd{E_\text{inj}},
\end{equation}
$Q$~being the injection spectrum (number of particles emitted per unit energy per unit time) of each source.
Likewise, we can define~$\mathcal{Q}(E_\text{inj})=n_\text{s} Q(E_\text{inj})$.

In the cases where the injection spectrum and density of sources depends on energy and possibly time (or equivalently redshift) but not on position, e.g.~$\mathcal{Q} = \mathcal{Q}(E_\text{inj},t)$, it can be shown that the expected fluxes at Earth are given by
\begin{equation}
J_i(E_\text{Earth}) =\frac{c}{4\pi} \int_{t_{\min}}^{t_0} \int_0^{+\infty} T_{ij}(E_\text{Earth}|\Einj, t) \mathcal{Q}_j(\Einj,t) \dd{\Einj} \dd{t},
\end{equation}
where $T_{ij}(E_\text{Earth}|\Einj, t)$ is the average number of particles of type~$i$ with energy~$E_\text{Earth}$ at the present time~$t_0$ from each particle of type~$j$ injected with energy~$\Einj$ at time~$t$. Therefore, in the case~$\mathcal{Q}(E_\text{inj},t) = \mathcal{Q}_0(E_\text{inj})S(z)$, in order to correctly analyse simulations of UHECR propagation in which the source positions are sampled from a uniform distribution in~$z$, each event must be weighed by a factor proportional to~$S(z)\abs{\dd{t}/\dd{z}}$.

\section{Photodisintegration models}\label{sec:disimodels}
Measurements of photodisintegration cross sections are not available for all nuclides, and when there are, sometimes only the total and/or the one-neutron ejection cross sections have been measured. Cross sections for exclusive channels in which charged fragments are ejected are hard to measure, because such ejectiles tend to undergo multiple scattering in the target. Various phenomenological models have been used in UHECR propagation studies to treat these processes.

The model by Puget, Stecker and Bredekamp (1976)~\cite{bib:PSBsigma}\index{PSB cross sections}
does not distinguish between protons and neutrons,
treating only beta-decay stable isobars for each mass number~$2\le A \le 4$ and~$9\le A \le 56$ (51~nuclides in total).
For each such nuclide, three types of photodisintegration processes are modelled:
one- and two-nucleon ejection for $\epsilon_{\min} = 2~\MeV < \epsilon' < \epsilon_1 = 30~\MeV$,
with cross sections approximated as truncated Gaussians
\begin{equation}
\sigma_i(\epsilon') = \xi_i\frac{\Sigma_\text{d}}{W_i}\exp(-2\frac{(\epsilon'-\epsilon_{0i})^2}{\Delta_i^2}), \quad i=1,2,
\label{eq:PSB12}
\end{equation}
where the peak position~$\epsilon_0$, normalized height~$\xi$ and width~$\Delta$ for each channel for each nuclide are listed in \TableRef{tab:PSBparams}, and $\Sigma_\text{d}$~and $W_i$~are normalization constants
\begin{align}
\Sigma_\text{d} &= 60\frac{NZ}{A}~\MeV~\mb, & W_i &= \int_{\epsilon_{\min}}^{\epsilon_1}\exp(-2\frac{(\epsilon'-\epsilon_{0i})^2}{\Delta_i^2})\dd{\epsilon'},
\end{align}
$N$~and $Z$~being the number of neutrons and protons respectively; and
multi-nucleon ejection for $\epsilon_1 = 30~\MeV < \epsilon' < \epsilon_{\max} = 150~\MeV$,
with constant cross sections
\begin{equation}
\sigma_3(\epsilon') = \frac{\zeta\Sigma_\text{d}}{\epsilon_{\max} - \epsilon_1}
\label{eq:PSB3}
\end{equation}
and fixed branching ratios for the number of nucleons ejected, listed in \TableRef{tab:PSBbranch}.
All other processes, e.g.~those in which deuterons or alpha particles are ejected,
are neglected in this model. An exception is beryllium-9, for which the only process modelled
is fragmentation into one nucleon and two alpha particles.

This model was refined by Stecker and Salamon (1999)~\cite{bib:SSthresholds}\index{Stecker--Salamon thresholds},
by replacing the $2~\MeV$~threshold with the actual kinematic threshold~$\epsilon_{\min}$ for one-
and two-nucleon ejection processes, also listed in \TableRef{tab:PSBparams}.

\begin{longtable}[c]{cc|cccc|cccc|c}
  ~ & ~ & \multicolumn{4}{c}{one-nucleon} & \multicolumn{4}{c}{two-nucleon} & ~ \\
  $A$ & $Z$ & $\epsilon_{\min}$ & $\epsilon_0$ & $\xi$ & $\Delta$ & $\epsilon_{\min}$ & $\epsilon_0$ & $\xi$ & $\Delta$ & $\zeta$ \\
  \hline
$56$ & $26$ & $10.2$ & $18$ & $0.98$ & $\phantom{0}8$ & $18.3$ & $22$ & $0.15$ & $\phantom{0}7$ & $0.95$ \\
$55$ & $25$ & $\phantom{0}8.1$ & $18$ & $0.93$ & $\phantom{0}7$ & $17.8$ & $24$ & $0.20$ & $\phantom{0}8$ & $0.95$ \\
$54$ & $26$ & $\phantom{0}8.9$ & $18$ & $0.93$ & $\phantom{0}7$ & $15.4$ & $24$ & $0.20$ & $\phantom{0}8$ & $0.95$ \\
$53$ & $24$ & $\phantom{0}7.9$ & $18$ & $1.03$ & $\phantom{0}7$ & $18.4$ & $24$ & $0.10$ & $\phantom{0}8$ & $0.95$ \\
$52$ & $24$ & $10.5$ & $18$ & $1.08$ & $\phantom{0}7$ & $18.6$ & $24$ & $0.05$ & $\phantom{0}8$ & $0.95$ \\
$51$ & $23$ & $\phantom{0}8.1$ & $19$ & $1.02$ & $\phantom{0}7$ & $19.0$ & $25$ & $0.11$ & $\phantom{0}6$ & $0.95$ \\
$50$ & $24$ & $\phantom{0}9.6$ & $19$ & $1.03$ & $\phantom{0}8$ & $16.3$ & $25$ & $0.10$ & $\phantom{0}6$ & $0.95$ \\
$49$ & $22$ & $\phantom{0}8.1$ & $19$ & $1.03$ & $\phantom{0}8$ & $19.6$ & $25$ & $0.10$ & $\phantom{0}6$ & $0.95$ \\
$48$ & $22$ & $11.4$ & $19$ & $1.03$ & $\phantom{0}8$ & $19.9$ & $25$ & $0.10$ & $\phantom{0}6$ & $0.95$ \\
$47$ & $22$ & $\phantom{0}8.9$ & $19$ & $1.03$ & $\phantom{0}8$ & $18.7$ & $25$ & $0.10$ & $\phantom{0}6$ & $0.95$ \\
$46$ & $22$ & $10.3$ & $19$ & $1.03$ & $\phantom{0}8$ & $17.2$ & $25$ & $0.10$ & $\phantom{0}6$ & $0.95$ \\
$45$ & $21$ & $\phantom{0}6.9$ & $19$ & $0.97$ & $\phantom{0}9$ & $18.0$ & $26$ & $0.15$ & $\phantom{0}8$ & $0.95$ \\
$44$ & $20$ & $11.1$ & $20$ & $0.92$ & $\phantom{0}9$ & $19.1$ & $26$ & $0.20$ & $\phantom{0}8$ & $0.96$ \\
$43$ & $20$ & $\phantom{0}7.9$ & $20$ & $0.97$ & $\phantom{0}8$ & $18.2$ & $26$ & $0.15$ & $\phantom{0}8$ & $0.96$ \\
$42$ & $20$ & $10.3$ & $20$ & $1.02$ & $\phantom{0}7$ & $18.1$ & $26$ & $0.10$ & $\phantom{0}8$ & $0.96$ \\
$41$ & $19$ & $\phantom{0}7.8$ & $20$ & $0.92$ & $\phantom{0}6$ & $17.7$ & $26$ & $0.20$ & $\phantom{0}8$ & $0.96$ \\
$40$ & $20$ & $\phantom{0}8.3$ & $20$ & $0.84$ & $\phantom{0}6$ & $14.7$ & $26$ & $0.28$ & $10$ & $0.96$ \\
$39$ & $19$ & $\phantom{0}6.4$ & $20$ & $0.73$ & $\phantom{0}7$ & $16.6$ & $25$ & $0.38$ & $12$ & $0.98$ \\
$38$ & $18$ & $10.2$ & $18$ & $0.86$ & $\phantom{0}8$ & $18.6$ & $22$ & $0.24$ & $\phantom{0}8$ & $0.98$ \\
$37$ & $17$ & $\phantom{0}8.4$ & $20$ & $0.81$ & $\phantom{0}7$ & $18.3$ & $24$ & $0.28$ & $\phantom{0}7$ & $1.00$ \\
$36$ & $18$ & $\phantom{0}8.5$ & $22$ & $0.82$ & $12$ & $14.9$ & $22$ & $0.25$ & $12$ & $1.00$ \\
$35$ & $17$ & $\phantom{0}6.4$ & $20$ & $0.87$ & $\phantom{0}7$ & $17.3$ & $26$ & $0.22$ & $10$ & $1.00$ \\
$34$ & $16$ & $10.9$ & $22$ & $0.87$ & $12$ & $20.1$ & $22$ & $0.20$ & $12$ & $1.00$ \\
$33$ & $16$ & $\phantom{0}8.6$ & $22$ & $0.82$ & $12$ & $17.5$ & $22$ & $0.25$ & $12$ & $1.00$ \\
$32$ & $16$ & $\phantom{0}8.9$ & $22$ & $0.97$ & $12$ & $16.2$ & $30$ & $0.10$ & $12$ & $1.00$ \\
$31$ & $15$ & $\phantom{0}7.3$ & $21$ & $0.85$ & $\phantom{0}8$ & $17.9$ & $29$ & $0.20$ & $12$ & $1.02$ \\
$30$ & $14$ & $10.6$ & $20$ & $0.83$ & $\phantom{0}7$ & $19.1$ & $26$ & $0.20$ & $\phantom{0}8$ & $1.04$ \\
$29$ & $14$ & $\phantom{0}8.5$ & $20$ & $0.83$ & $\phantom{0}7$ & $20.1$ & $26$ & $0.20$ & $\phantom{0}8$ & $1.04$ \\
$28$ & $14$ & $11.6$ & $21$ & $1.01$ & $\phantom{0}8$ & $19.9$ & $30$ & $0.02$ & $\phantom{0}8$ & $1.04$ \\
$27$ & $13$ & $\phantom{0}8.3$ & $21$ & $0.80$ & $\phantom{0}8$ & $19.4$ & $29$ & $0.20$ & $12$ & $1.05$ \\
$26$ & $12$ & $11.1$ & $18$ & $0.77$ & $\phantom{0}8$ & $18.4$ & $26$ & $0.20$ & $\phantom{0}8$ & $1.08$ \\
$25$ & $12$ & $\phantom{0}7.3$ & $23$ & $0.77$ & $\phantom{0}9$ & $19.0$ & $28$ & $0.20$ & $\phantom{0}7$ & $1.08$ \\
$24$ & $12$ & $11.7$ & $19$ & $0.94$ & $11$ & $20.5$ & $29$ & $0.03$ & $\phantom{0}6$ & $1.08$ \\
$23$ & $11$ & $\phantom{0}8.8$ & $22$ & $0.83$ & $12$ & $19.2$ & $25$ & $0.12$ & $10$ & $1.09$ \\
$22$ & $10$ & $10.4$ & $22$ & $0.81$ & $12$ & $17.1$ & $21$ & $0.11$ & $\phantom{0}4$ & $1.09$ \\
$21$ & $10$ & $\phantom{0}6.8$ & $22$ & $0.84$ & $12$ & $19.6$ & $25$ & $0.08$ & $\phantom{0}6$ & $1.09$ \\
$20$ & $10$ & $12.8$ & $22$ & $0.87$ & $12$ & $20.8$ & $26$ & $0.05$ & $\phantom{0}8$ & $1.09$ \\
$19$ & $\phantom{0}9$ & $\phantom{0}8.0$ & $23$ & $0.76$ & $14$ & $16.0$ & $29$ & $0.14$ & $14$ & $1.10$ \\
$18$ & $\phantom{0}8$ & $\phantom{0}8.0$ & $24$ & $0.67$ & $\phantom{0}9$ & $12.2$ & $29$ & $0.20$ & $10$ & $1.10$ \\
$17$ & $\phantom{0}8$ & $\phantom{0}4.1$ & $24$ & $0.77$ & $\phantom{0}9$ & $16.3$ & $29$ & $0.20$ & $10$ & $1.10$ \\
$16$ & $\phantom{0}8$ & $12.1$ & $24$ & $0.83$ & $\phantom{0}9$ & $22.3$ & $30$ & $0.04$ & $10$ & $1.10$ \\
$15$ & $\phantom{0}7$ & $10.2$ & $23$ & $0.73$ & $10$ & $18.4$ & $23$ & $0.10$ & $10$ & $1.07$ \\
$14$ & $\phantom{0}7$ & $\phantom{0}7.6$ & $23$ & $0.46$ & $10$ & $12.5$ & $23$ & $0.37$ & $10$ & $1.07$ \\
$13$ & $\phantom{0}6$ & $\phantom{0}4.9$ & $23$ & $0.71$ & $\phantom{0}8$ & $20.9$ & $27$ & $0.05$ & $\phantom{0}8$ & $1.06$ \\
$12$ & $\phantom{0}6$ & $16.0$ & $23$ & $0.76$ & $\phantom{0}6$ & $27.2$ & $27$ & $0.00$ & $\phantom{0}8$ & $1.06$ \\
$11$ & $\phantom{0}5$ & $11.2$ & $26$ & $0.85$ & $11$ & $18.0$ & $26$ & $0.15$ & $11$ & $1.03$ \\
$10$ & $\phantom{0}5$ & $\phantom{0}6.6$ & $25$ & $0.54$ & $11$ & $\phantom{0}8.3$ & $25$ & $0.15$ & $11$ & $1.03$ \\
$\phantom{0}9$ & $\phantom{0}4$ & $\phantom{0}1.7$ & $26$ & $0.67$ & $20$ & $18.9$ & $25$ & $0.00$ & $11$ & $1.00$ \\
$\phantom{0}4$ & $\phantom{0}2$ & $19.8$ & $27$ & $0.47$ & $12$ & $26.1$ & $45$ & $0.11$ & $40$ & $1.11$ \\
$\phantom{0}3$ & $\phantom{0}2$ & $\phantom{0}5.5$ & $13$ & $0.33$ & $18$ & $\phantom{0}7.7$ & $15$ & $0.33$ & $13$ & $1.11$ \\
$\phantom{0}2$ & $\phantom{0}1$ & $\phantom{0}2.2$ & $\phantom{0}5$ & $0.97$ & $\phantom{0}9$ & $\phantom{0}2.2$ & $15$ & $0.00$ & $13$ & $0.00$ \\
  \caption[PSB cross section parameters]{Parameters of PSB cross sections with Stecker and Salamon thresholds (all energies in MeV)}
  \label{tab:PSBparams}
\end{longtable}

\begin{table}
  \centering
  \begin{tabular}{r|ccccc cccc }
  $A$ & 1 & 2 & 3 & \multicolumn{6}{l}{etc.}\\
  \hline
  3--4 &   $\phantom{0}80\%$ & $20\%$ & \multicolumn{7}{l}{~}\\
  9 &                $100\%$ & \multicolumn{8}{l}{~} \\
  10--22 & $\phantom{0}10\%$ & $30\%$ & $10\%$ &           $10\%$ & $20\%$ & $20\%\phantom{.0}$ & \multicolumn{3}{l}{~} \\
  23--56 & $\phantom{0}10\%$ & $35\%$ & $10\%$ & $\phantom{0}5\%$ & $15\%$ & $\phantom{0}4.5\%$ &
           $4.0\%$ & $3.5\%$ & $3.0\%$  \\ (cont.) & $2.5\%$ & $2.0\%$ & $1.8\%$ & $1.5\%$ & $1.2\%$ & $1.0\%$ \\
  \end{tabular}
  \caption[PSB cross section branching ratios]{Branching ratios for the number of nucleons ejected in the PSB model for~$30~\MeV < \epsilon' < 150~\MeV$ as a function of the parent nucleus mass number}
  \label{tab:PSBbranch}
\end{table}

A more complete model is TALYS~\cite{bib:TALYS}\index{TALYS}, a program that can simulate nuclear reactions for a variety
of projectile types and a wide range of projectile energies, computing cross sections for all exclusive
channels, $\sigma_{n_\n n_\p n_{\rm d} n_{\rm t} n_{\rm h} n_\alpha}$~being the cross section for the channel
in which $n_\n$~neutrons, $n_\p$~protons, $n_{\rm d}$~deuterons, $n_{\rm t}$~tritium nuclei, $n_{\rm h}$~helium-3 nuclei
and $n_\alpha$~helium-4 nuclei are ejected.

As discussed in Ref.~\cite{bib:SALpropa}, released versions of TALYS have been found to be in worse
agreement with the measured data for total photodisintegration cross sections for photon energies and mass numbers relevant for UHECR propagation
than the preliminary version used in Ref.~\cite{bib:Khan} when used with their default settings. The differences include the default use in the released versions of TALYS of the Brink--Axel Lorentzian model for the $E1$~gamma-ray strength function, whereas in Ref.~\cite{bib:Khan} the Kopecky--Uhl generalized Lorenzian model was used, and the default use in the released versions of TALYS of GDR parameters from the RIPL-2 database~\cite{bib:RIPL2}, whereas in Ref.~\cite{bib:Khan} values from the IAEA atlas~\cite{bib:IAEAatlas}, listed in \TableRef{tab:GDRparams}, were used. For this reason, TALYS-1.6 with parameters restored\index{TALYS ``restored''} to the values used in~Ref.~\cite{bib:Khan} was used to compute the parameters in  \nolinkurl{pars_talysfixed.txt}.
Also, all versions of TALYS largely overestimate the cross sections for channels in which alpha particles are emitted for the few nuclides for which measured data for these channels are available; these channels are neglected altogether in the PSB model. (The choice of $E1$~strength function model and GDR parameters in TALYS only affects the total cross sections but not the branching ratios.)

\begin{table}
\centering
\begin{tabular}{cc|ccc|ccc|c}
    $A$ & $Z$ & ${E_0}$ & ${\sigma_0}$ & $\Gamma_0$ & $E_1$ & $\sigma_1$ & $\Gamma_1$ & source \\
    \hline
    12 & \phantom{0}6  &  22.70  &  21.36  &  \phantom{0}6.00  &  &  &  &  atlas~\cite{bib:IAEAatlas} \\
    14 & \phantom{0}7  &  22.50  &  27.00  &  \phantom{0}7.00  &  &  &  &  atlas \\
    16 & \phantom{0}8  &  22.35  &  30.91  &  \phantom{0}6.00  &  &  &  &  atlas \\
    23 & 11 &  23.00  &  15.00  & 16.00  &  &  &  &  atlas \\
    24 & 12 &  20.80  &  41.60  &  \phantom{0}9.00  &  &  &  &  atlas \\
    27 & 13 &  21.10  &  12.50  &  \phantom{0}6.10  &  29.50  &  \phantom{0}6.70  &  8.70 &  RIPL-2~\cite{bib:RIPL2} \\
    28 & 14 &  20.24  &  58.73  &  \phantom{0}5.00  &  &  &  &  atlas \\
    40 & 18 &  20.90  &  50.00  & 10.00  &  &  &  &  atlas \\
    40 & 20 &  19.77  &  97.06  &  \phantom{0}5.00  &  &  &  &  atlas \\
    51 & 23 &  17.93  &  53.30  &  \phantom{0}3.62  &  20.95  &  40.70  &  7.15 &  RIPL-2 \\
    55 & 25 &  16.82  &  51.40  &  \phantom{0}4.33  &  20.09  &  45.20  &  4.09 &  RIPL-2 \\
\end{tabular}
\caption[GDR parameters used with TALYS ``restored'']{GDR parameters used with TALYS ``restored'' (all energies and areas in $\MeV$ and $\mb$ respectively) for the Kopecky--Uhl generalized Lorentzian model of the $E1$ strength function; for nuclides not listed here and for higher-order contributions, formulas described in the TALYS-1.6 user manual are used.}
\label{tab:GDRparams}
\end{table}

\bibliographystyle{JHEP}
\bibliography{simprop}

\end{document}